\documentclass[manuscript]{aastex}

\shorttitle{ChromaStarPy}
\shortauthors{Short, Bayer, \& Burns}

\begin{document}


\title{ChromaStarPy: A stellar atmosphere and spectrum modeling and visualization lab in python}


\author{C. Ian Short}
\affil{Department of Astronomy \& Physics and Institute for Computational Astrophysics, Saint Mary's University,
    Halifax, NS, Canada, B3H 3C3}
\email{ian.short@smu.ca}

\author{Jason H. T. Bayer}
\affil{Department of Astronomy \& Physics and Institute for Computational Astrophysics, Saint Mary's University,
    Halifax, NS, Canada, B3H 3C3}

\author{Lindsey M. Burns}
\affil{Department of Astronomy \& Physics and Institute for Computational Astrophysics, Saint Mary's University,
    Halifax, NS, Canada, B3H 3C3}




\begin{abstract}

We announce ChromaStarPy, 
an integrated general stellar atmospheric modeling and
spectrum synthesis code written entirely in python V. 3.  
ChromaStarPy is a direct port of the ChromaStarServer (CSServ) Java modeling code
described in earlier papers in this series, and many of the associated JavaScript (JS) post-processing procedures
have been ported and incorporated
into CSPy so that students have access to ready-made ``data products''.
A python integrated development environment (IDE) allows a student in a more advanced course  
to experiment with the code and to graphically visualize intermediate and final results, {\it ad hoc},
as they are running it.  CSPy allows
students and researchers to compare modeled to observed spectra in the same IDE in which they are 
processing observational data, while having complete control over the stellar parameters 
affecting the synthetic spectra. 
We also take the opportunity to describe improvements that have been made to 
the related codes, ChromaStar (CS), CSServ and ChromaStarDB (CSDB) that,
where relevant, have also been incorporated into CSPy.  
The application may be found at the home page of the OpenStars project: www.ap.smu.ca/OpenStars/. 
   
\end{abstract}



\section{Introduction}

  A refreshing recent development in astronomy research and higher education 
 is the advent of computational tools for commonplace platforms 
of the type that students are likely to own.  This has been enabled by the 
power and capacity of consumer devices, which now exceed those of institutional workstations of
previous decades.  This makes it easier to assign lab activities as coursework, 
and for motivated students to explore computational tools on their own initiative. 

\paragraph{}

The development of the python programing language and of free multi-platform python distributions
that include an integrated development environment (IDE), such as 
spyder, and plotting libraries such as matplotlib \citep{matplotlib}, has been consequential.  The IDE 
provides an interactive 
environment for code analysis and development, and a visualization environment similar to
that of more established scientific plotting packages.  
Python is increasingly being used by the observational astronomy community, and
there are standard libraries of astronomical procedures such as astropy  \citep{astropy} and 
PyRAF (see, for example, \citet{pyraf}).  
An added incentive for python deployments is the advent of ``notebook'' applications,
such as juPyter, that are equipped with python kernels, and that allow code blocks to 
be interspersed with marked-up text and plots. 
The iSpec spectral analysis code \citep{ispec}
is a recent example of an institutional research capability being made available for more commonly
accessible platforms by implementation in python.      

%
 
\paragraph{}

 We introduce ChromaStarPy (CSPy, \dataset{DOI: zenodo.1095687}), a general 1D static plane-parallel LTE atmospheric modeling, 
spectrum synthesis, and post-processing code written entirely in python, equipped with an optional jupyter notebook
interface, and freely available under the MIT license as a source tarball from the home page of the OpenStars project
(www.ap.smu.ca/OpenStars).  
Additionally, there is a GitHub repository for version-controlled collaborative development
(https://github.com/sevenian3/ChromaStarPy), and we note that the version numbering scheme is date-based 
using the ISO 8601 system (YYYY-MM-DD). 
The code is a direct port of the ChromaStarServer (CSServ) Java application
that is described in detail in \citet{gss16} (S16) and \citet{csdb17} (S17), and its methods and 
procedures are the same as that code.  Like CSServ, CSPy adopts significant modeling approximations to
expedite the computation so as to be more suitable for pedagogical application, while retaining a level 
of modeling realism suitable for projects in a pedagogical context, and for research projects of a 
carefully limited scope.  Given the significance 
of a port to a more scientific
programing language, we take the opportunity to review the methods and major approximations being
made in Section \ref{sMethods}, present comparisons with observations in Section \ref{sTests}, 
and in Section \ref{sApps} we propose sample activities for 
which CSPy is especially suitable.  Related applications that have been described elsewhere include
ChromaStarDB (CSDB), a version of CSServ that implements the line list as an SQL database
and allows for novel flexibility in selecting or deselecting which spectral lines to include in a spectrum synthesis
based on a wide variety of selection criteria \citep{csdb17}, and
ChromaStar (CS), a pure JavaScript and html version of the code that uses a small line list of only ~20
lines that runs entirely in the client browser and is suitable for broader education and public outreach (EPO)
\citep{pedagogy}.
In Section \ref{sDev} we describe recent improvements to all
these related OpenStars codes that 
have been made since the last published report on those codes, and which, where relevant, are 
also reflected in CSPy. 

\section{Methods \label{sMethods}}

\subsection{Atmospheric structure and spectrum modeling}

A detailed description of the atmospheric structure modeling and the spectrum synthesis performed 
by CSServ, and ported to CSPy, including the crucial 
approximations for expediting the calculation, along with the justifications and discussion of the 
limitations, are to be found in \citet{gss16} and \citet{csdb17}.
The most significant distinctions from the simplest research-grade modeling are that CSPy 
obtains the vertical kinetic temperature structure, $T_{\rm kin}(\tau)$, as a function of optical depth,
 $\tau$, by simple re-scaling with effective temperature, $T_{\rm eff}$, from one of three research-grade 
template models that sample the populated quadrants of the HR diagram, and that the Voigt profiles
of spectral lines are approximated with a series expansion \citep{gray}.

CSPy automatically performs a spectrum synthesis calculation after computing the atmospheric structure
every time it is run.  The code can be run in a ``spectrum synthesis'' mode in which a previously
computed structure is read from a imported source file (a python module), in which case CSPy
automatically limits itself to one iteration of the structure to ensure that it has all needed
quantities and proceeds quickly to the spectrum synthesis stage.  This expedites activities
for which the spectrum synthesis parameters should be varied for a fixed atmospheric structure.

\subsubsection{Line lists}

As of this work CSPy uses a line list of $\sim 26\,000$ atomic lines in the wavelength ($\lambda$) range from 260 to 2600 nm 
(covering the Johnson $U$, $B$, $V$, $R$, $I$, $J$, $H$, and $K$ bands) extracted from the NIST Atomic Spectra
Database \citep{nist}.  CSPy expects to read the line list in byte-data form and the distribution includes both the
original ascii and the byte-data versions of the line list.  
The utility procedure LineListPy.py is included with the distribution and converts an ascii line list in the
format produced by the Atomic Spectra Database into the byte-data format expected by CSPy.  This allows a 
more advanced user to produce their own custom line lists.  Care must be taken to set the numerous output options
in the Atomic Spectra Database interface so as to produce output with fields and units that match what 
LineListPy.py expects, and some manual post-processing of the ascii output from the database may be necessary to remove
special characters.  The ascii version of the line list provided with the distribution serves as a template
for what is required.

\subsection{User-defined two-level atom \label{stla}}

A pedagogically significant additional capability that has been ported from CS is modeling of the energy level 
populations and spectral line profile of an artificial atomic species with only two abound states - a two 
level atom (TLA).  The user can adjust the abundance of the TLA species on the logarithmic ``$A_{12}$'' scale 
used in stellar spectroscopy, and the atomic and line transition data for the TLA.  This includes the
excitation energy of the lower $E$-level ($E_{\rm i}$), the central wavelength, $\lambda_{\rm o}$, of the spectral line 
(which, along with the value of $E_{\rm i}$ determines the value of the upper energy level, $E_{\rm j}$),
and the oscillator strength, $f$, among other things.  The user can ascribe the TLA to any of the first four ionization stages of
the corresponding artificial ``element'', and set the first four ionization energies.  The user can
adjust a pair of ``fudge'' factors - one for the overall level of the background continuous opacity 
($\kappa^{\rm c}_\lambda$) (this also affects the overall spectrum synthesis), and the logarithmic enhancement of 
the Lorentzian wings in the case that the TLA line is saturated.  The macroscopic broadening parameters
accounting for macro-turbulence and rotation are also applied to the TLA line.  CSPy computes the 
equivalent width, $W_\lambda$, of the TLA line.   

\subsection{Limb darkening coefficients \label{sldc}}

Like CS, CSServ, and CSDB, CSPy computes monochromatic continuum linear limb darkening coefficients (LDCs),
$\epsilon(\lambda)$, at each of the $\lambda$ points sampling the continuum $I^{\rm c}_\lambda(\lambda, \cos\theta)$ distribution,
where $\theta$ is the angle of emergence of an $I_\lambda$ pencil-beam with respect to the local surface normal, and
 $\epsilon$ is defined by the linear limb darkening law,

\begin{equation}
I^{\rm c}_\lambda(\cos\theta)/I^{\rm c}_\lambda(\cos\theta=1) = (1-\epsilon) + \epsilon\cos\theta
\label{ldceq}
\end{equation}

  Currently, CSPy simply solves Eq. \ref{ldceq} for $\epsilon$ separately for each $(\lambda, \cos\theta)$ pair
and averages the resulting $\epsilon(\cos\theta, \lambda)$ values over $\cos\theta$ at each $\lambda$ value to
produce $\epsilon(\lambda$) values.  We plan to eventually replace this with a procedure that finds the best fitting
$\epsilon$ value for the $I^{\rm c}_\lambda(\cos\theta)/I^{\rm c}_\lambda(\cos\theta=1)$ values at each
$\lambda$ value.

\subsection{Comparison to iSpec}

iSpec is another astronomical tool for working with model stellar atmospheres and spectra that has been developed in python 
(see \citep{ispec}), and here we contrast it with CSPy.  The iSpec application is an environment that allows a user to make 
use of libraries of
atmospheric models pre-computed with research-grade modeling codes, and to compute spectra with their choice of
spectrum synthesis codes pre-compiled from Fortran.  This provides a powerful spectral modeling and analysis environment
for extracting information from spectra with underlying research-grade modeling codes.  By contrast, CSPy is an
atmospheric modeling and spectrum synthesis code written entirely in python, and could be run alongside iSpec in
a python IDE to generate the synthetic spectra.     

\section{Tests \label{sTests}}

\begin{figure}
\plotone{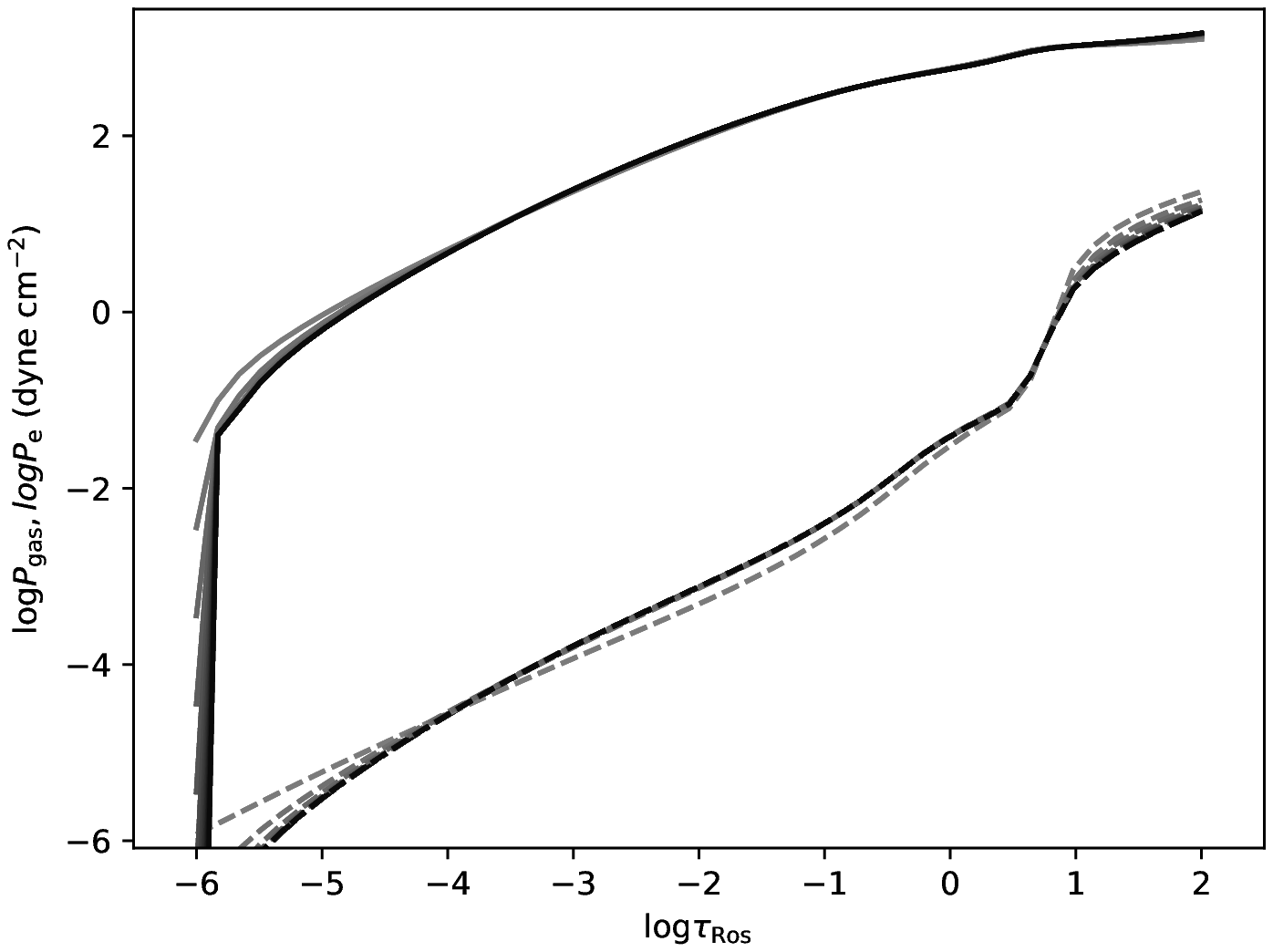}
\caption{The $\log P_{\rm gas}(\log\tau)$ (solid lines) and $\log P_{\rm e}(\log\tau)$ (dashed lines) structures after 
each of 12 iterations 
of the atmospheric structure of a model of $T_{\rm eff}/\log g/[{\rm Fe}/{\rm H}]$ equal to 3600 K/0.0/0.0.  
Darker lines indicate later iterations.  The initial guess consisted of $\log P_{\rm gas}(\log\tau)$ and $\log P_{\rm e}(\log\tau)$ 
structures that were approximately re-scaled with $T_{\rm eff}$, $\log g$ and $[{\rm Fe}/{\rm H}]$ from a template model of
4250 K/2.0/0.0 (see text). 
  \label{fConverge}
}
\end{figure}

\begin{figure}
\plotone{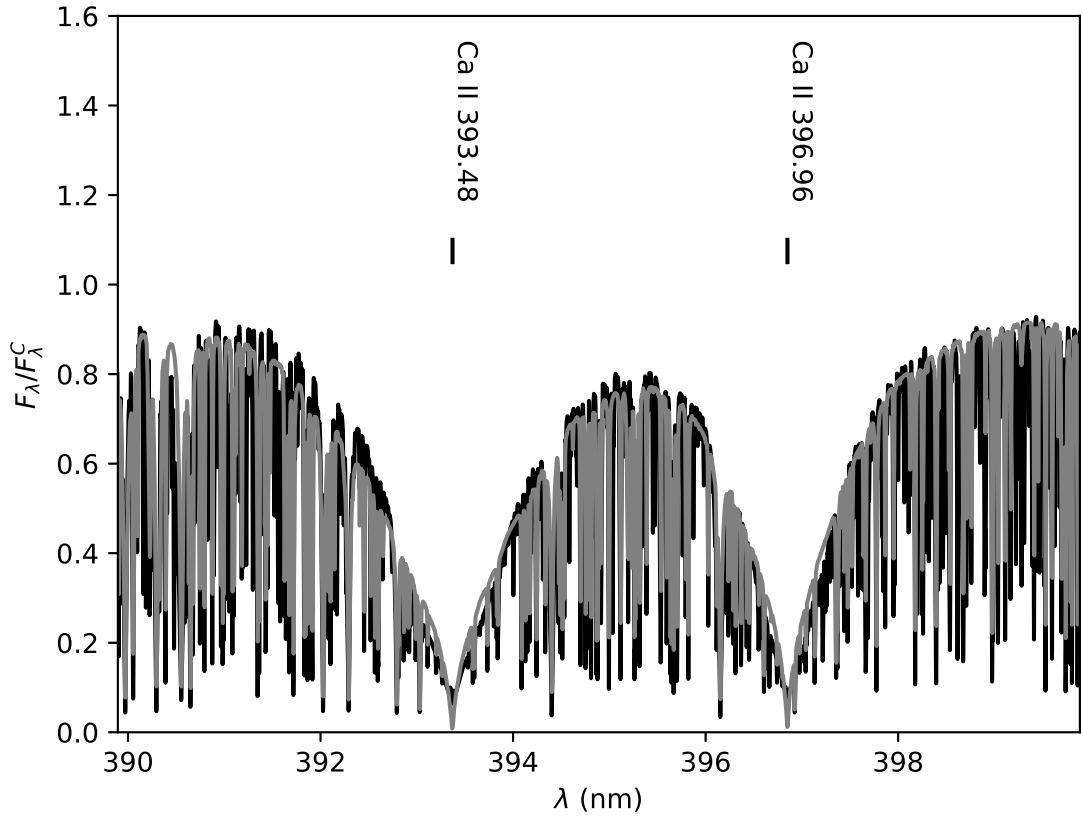}
\caption{The observed \ion{Ca}{2} HK region of the solar flux spectrum as presented by \citet{solarflux} (black line) and as computed with 
CSPy (gray line) with a Lorentzian broadening enhancement factor of $10^{0.5}$. 
  \label{fSunCaIIHK}
}
\end{figure}

\begin{figure}
\plotone{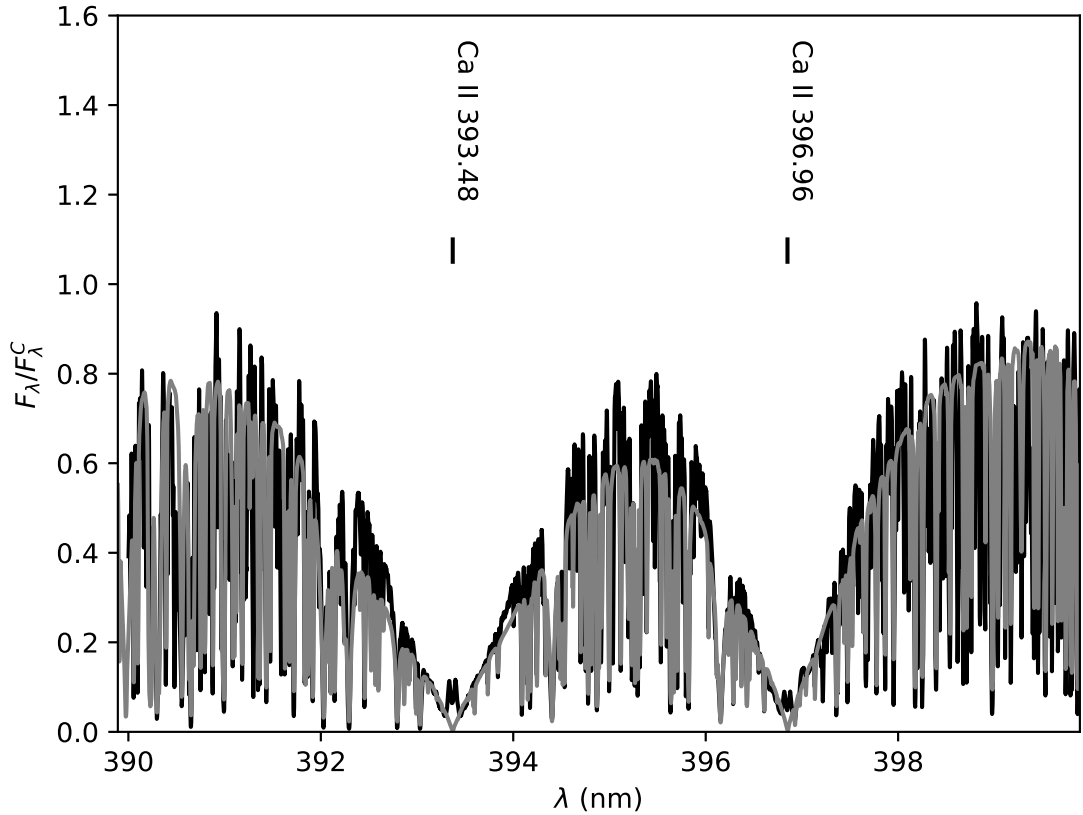}
\caption{The \ion{Ca}{2} HK region of the flux spectrum of Arcturus ($\alpha$ Boo) as observed by \citet{arcturusFlux} (black line) 
and as computed with CSPy (gray line) with no Lorentzian broadening enhancement.  We note that 
our models do not have a chromospheric temperature inversion and we do not
expect to fit the core emission reversals in the strong HK lines. 
  \label{fAlfBooCaIIHK}
}
\end{figure}

\begin{figure}
\plotone{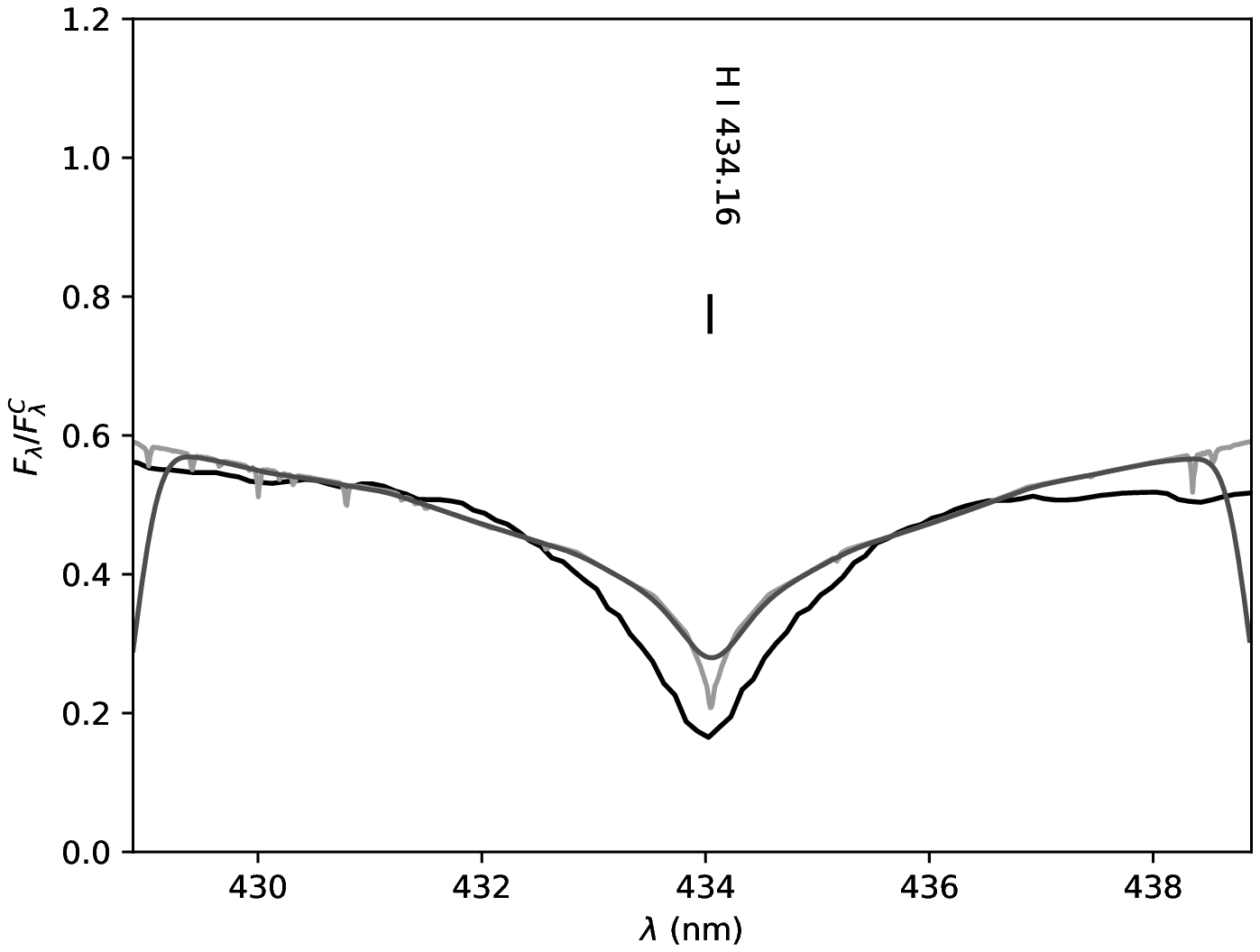}
\caption{The \ion{H}{1} Balmer $\gamma$ region of the flux spectrum of Vega ($\alpha$ Lyr) as observed by \citet{vegaFlux} (black line) 
and as computed with CSPy (light gray line), and then instrumentally broadened to $R = 2000$ (dark gray line).  We note that 
we expect to fit the far wings of the linear Stark broadening profile with our
current treatment. 
  \label{fVegaHgamma}
}
\end{figure}

   Fig. \ref{fConverge} shows the $\log P_{\rm gas}(\log\tau)$ and $\log P_{\rm e}(\log\tau)$ structures after each of 12 iterations 
of the structure equations for a red supergiant model of $T_{\rm eff}/\log g/[{\rm Fe}/{\rm H}]$ equal to 3600 K/0.0/0.0.  For these input
parameters, CSPy determined its initial guess at these structures by approximately re-scaling from the red giant
template of Section \ref{sMethods} (4250 K/2.0/0.0).  In addition to demonstrating the convergence properties of 
our procedure, Fig. \ref{fConverge} is a good example of the kind of plot that a student can easily make in a python IDE
on their own device by simply instrumenting the code with matplotlib plot statements.   

  \paragraph{}

  S16 presented comparisons of synthetic spectra computed with CSServ to those computed with Phoenix V. 15 \citep{phoenix} for the \ion{Ca}{2} 
region for stars of $T_{\rm eff}$/$\log g$/$[{{\rm Fe}\over {\rm H}}]$/$\xi_{\rm T}$ equal to 5000 K/4.5/0.0/1.0 km s$^{\rm -1}$
and 5000 K/2.5/0.0/1.0 km s$^{\rm -1}$, and the \ion{Mg}{2} $\lambda 4481$ region for
a model of 10000 K/4.0/0.0/1.0 km s$^{\rm -1}$.  As of S16, we were not yet ready to compare synthetic and observed spectra
in the vicinity of \ion{H}{1} lines for early-type stars because we had not yet incorporated linear Stark broadening for the 
Balmer lines, and the \ion{Mg}{2} $\lambda 4481$ line is the next most important MK classification diagnostic for these stars.  
S17 presented the equivalent comparisons for
the TiO $C^3\Delta - X^3\Delta$ ($\alpha$ system, band origin, $\lambda_{00}=517.02$ nm)
and  $c^1\Phi - a^1\Delta$ ($\beta$ system, $\lambda_{00}=560.52$ nm) bands for stars of solar metallicity 
of $T_{\rm eff} = 3750$ K and $\log g$ values of 4.5 and 2.0, and of $T_{\rm eff} = 4250$ K and $\log g = 2.0$.  

\paragraph{}

  Fig. \ref{fSunCaIIHK} shows the flux spectrum, $F_\lambda$, of the Sun in the heavily blanketed \ion{Ca}{2} HK region 
based on observations with a resolving power, $R$, of $~300\, 000$ described by \citet{solarflux}, and as computed 
with CSPy with a Lorentzian broadening enhancement factor of $10^{0.5}$.  We note that the NIST atomic line list
is much less complete than current competitive research-grade line lists, and we do not expect to
completely treat the line blanketing in this region.  
We adopt the standard parameters of $T_{\rm eff}=5777$ K, $\log g=4.44$, and $[{\rm Fe}/{\rm H}]=0.0$, and a value of the microturbulent
velocity dispersion, $\xi_{\rm T}$, of 1 km s$^{-1}$.
Fig. \ref{fAlfBooCaIIHK} shows the same region for Arcturus ($\alpha$ Boo, \objectname{HD124897}, \objectname{HR5340}, K1-K1.5\,{\sc III}) 
as observed by \citet{arcturusFlux} at $R\approx 100\, 000$ and as computed with CSPy. 
We adopt the parameters of \citet{griffin_l99}, rounded to to the 
nearest canonical values, $T_{\rm eff}=4300$ K, $\log g=2.0$, and $[{\rm Fe}/{\rm H}]=-0.7$, with an $\alpha$-process element 
enhancement of $+0.3$ \citep{pdk93}, and adopt a value of $\xi_{\rm T}$ of 2 km s$^{-1}$.
Because our scaled radiative and convective thermal
equilibrium models lack a chromospheric temperature inversion, we do not expect to reproduce the emission core reversals
that are apparent in the observed spectrum of Arcturus.
Because of the high spectral resolution of
the observational material, and the approximate nature of the modeling that we are assessing, we make no attempt to 
convolve the synthetic spectra to match the instrumental resolution of the observations.

\paragraph{}

Fig. \ref{fVegaHgamma} presents a comparison of the \ion{H}{1} H$\gamma$ wings for Vega ($\alpha$ Lyr, HD 172167, HR 7001, A0 V)
as observed by \citet{vegaFlux} at $R\approx 2000$, and as computed with CSPy and broadened by convolution with a Gaussian kernel 
to the $R$ value of the observed spectrum.  H$\gamma$ is the longest wavelength Balmer line in the observed spectrum, and we chose it
for comparison because we expect it to be the least blended with neighboring lines. 
We adopt the parameters of \citet{castelli} of 
$T_{\rm eff}=9550$ K, $\log g=3.95$, $[{\rm Fe}/{\rm H}]=-0.5$, and $\xi_{\rm T}$ = 2 km s$^{-1}$ for the structure, and
values of $v_{\rm Rot}$ and $i$ of 275 km s$^{-1}$ and 5$^\circ$, respectively, corresponding to a $v\sin i$ value of 24 
km s$^{-1}$ \citep{vegaRot}, for the post-processing of the spectrum.  We note that 
currently we only treat
linear Stark broadening in the ``far'' wing, as described in S17, for $\Delta\lambda > 2\Delta\lambda_{\rm D}$, where $\Delta\lambda_{\rm D}$
is the Doppler width, and only expect to match the profile even approximately in that regime.

\section{Applications \label{sApps}}

  As an integrated stellar atmospheric modeling and spectrum synthesis code, CSPy takes as input
the standard parameters needed for un-blanketed, static, 1D, plane-parallel, LTE, scaled-solar abundance 
modeling: $T_{\rm eff}$, 
$\log g$, and $[{\rm Fe}/{\rm H}]$.  Because this is a pedagogical code, it also allows the user to specify a stellar mass, $M$, which
it uses to compute and display the stellar radius, $R$, and bolometric luminosity, $L_{\rm bol}$,
corresponding to the values of $T_{\rm eff}$ and $\log g$.  It also requires parameters specifying 
the spectrum synthesis: the wavelength range, $[\lambda_1, \lambda_2]$, the microturbulent
velocity dispersion, $\xi_{\rm T}$, and fudge factors for tuning the background continuum opacity,
$\kappa^{\rm c}_\lambda$, and the Lorentzian line broadening.  Because post-processing of the 
spectrum accounting for natural effects is integrated, it also allows specification
of the macroturbulent dispersion, $\xi_{\rm Macro}$, the surface equatorial rotation velocity,
$v_{\rm rot}$, and the inclination of the rotation axis to the line-of-sight, $i$. 

\paragraph{}

   Because python is an interpreted, rather than a compiled, language, it allows for flexible
diagnostic interaction when running in an IDE such as spyder.  The values of intermediate
variables can be inspected, {\it ad hoc}, at the console prompt.  When accompanied by a 
plotting facility, such as matplotlib, the code can be
instrumented with {\it ad hoc} plot statements that allow for visual inspection of how 
various structures are converging from one iteration to the next, as exemplified by Fig. 
\ref{fConverge}.  Students can edit
the code and rerun it while learning about, or developing, the code.  As a result, the
IDE is effectively an integrated computational astrophysics ``lab bench''.

\paragraph{}

  Students can post-process the synthetic spectra with operations such as convolution in the
IDE.  Students in observational astronomy courses who have acquired a sample observed stellar spectrum 
can import the observational data into the IDE and compare it, qualitatively or statistically, to 
suitably post-processed
model spectra generated {\it ad hoc} with various trial stellar and spectrum synthesis parameters.
Students can extract the $W_\lambda$ value of a spectral line,
either with the built-in procedure if they restrict the synthesis range to isolate one line, or
with a python procedure of their own.  
A local example is that students in the fourth year observational astronomy or experimental physics
courses in our program at Saint Mary's University can acquire CCD spectra with the spectrograph on the
60 cm telescope of the Burke-Gaffney Observatory (BGO), and compare model spectra
produced with CSPy in a python IDE.  

\subsection{TLA}

  The TLA (see section \ref{stla}) can be used to study the simple curve-of-growth (COG) of
a spectral line, which, here, is effectively being defined as $\log W_\lambda([{\rm A}/{\rm H}])$.  Because
the TLA is treated with a Voigt profile, the COG will exhibit the weak, strong, and saturated
regimes as $[{\rm A}/{\rm H}]$ increases.  The student can experimentally investigate how the COG
varies with $f$ and $E_{\rm i}$.  

\paragraph{}   

  The TLA can be used to model specific real spectral lines that are important diagnostics.
An effective example is to set the TLA parameters to those of the \ion{Ca}{1} $\lambda 4227$
line, and then the \ion{Ca}{2} K line, and to study the $\log W_\lambda(T_{\rm eff})$ 
behavior over the $T_{\rm eff}$ range of late-type stars (3500 - 6500 K).  This provides
a good example of the role that ionization equilibrium plays in the relation between 
$T_{\rm eff}$ and MK spectral class.

\section{Improvements to the OpenStars suite \label{sDev}}

   A number of significant improvements to the related codes CS, CSServ, and CSDB
have been made since our last report, and these have also been incorporated into CSPy
where relevant.

\subsection{Partition functions}

As of S17, CS, CSServ, and CSDB estimated the value of the partition function, $U_{\rm i}(\theta)$, for
species $i$ by linear interpolation among two values of $\theta$, 0.5 and 1.0, where $\theta\equiv 5040/T$,
with the $U_{\rm i}(\theta)$ values taken from \citet{allens}.  The treatment has been improved by 
interpolating in $T$ among the $U_{\rm i}(T)$ values of \citet{barklem} for $T$ values of 130, 500, 3000, 
8000, and 10000 K.  We incorporate the $T$ values below 3000 K in anticipation of eventually 
adapting the modeling so as to be more suitable for brown dwarfs.  For now, the interpolation in $T$ remains linear,
but this new treatment yields $n_{\rm i}(\tau)$ distributions that are more continuous and do not
suffer from the small discontinuities that were produced by the two-temperature $U_{\rm i}(\theta)$
treatment.

\subsection{Non-solar abundance distributions}

In addition to allowing for adjustment of the overall scaled-solar metallicity, $[{\rm Fe}/{\rm H}]$,
the codes now also allows the user to independently adjust the values of the logarithmic quantities
[{\rm He}/{\rm Fe}], [$\alpha$/{\rm Fe}], and [{\rm C}/{\rm O}], where $\alpha$ indicates eight $\alpha$-process elements
(O, Ne, Mg, Si, S, Ar, Ca, and Ti).  This affects the computation of the $P_{\rm gas}(\tau)$ and 
$P_{\rm e}(\tau)$ structures through the EOS and HSE treatment, and the value of the background
$\kappa_\lambda(\tau)$ distribution as well as having a direct effect on the strengths of the
relevant lines in 
the spectrum synthesis.  This allows the user to investigate the effect on both the atmospheric 
structure and spectrum of He enrichment in A and B stars, $\alpha$-enhancement in metal-poor
RGB stars, and enhanced C/O values in post-dredge-up AGB stars.  

\subsection{Photometry}

The Johnson $UBVRI$ filters employed by the integrated post-processing suite have now been supplemented 
with the $HJK$ filters, and we have adopted the
response curves of \citet{HJK}, as reported in the Asiago Database of Photometric Systems \citep{asiago}. 
The SED is now computed from 260 to 2600 nm, and the additional filters allow us to compute and
display the standard $V-K$ and $J-K$ color indices.  These are relevant as our modeling currently
extends down to $T_{\rm eff}$ values of 3400 K, and will become increasingly important
as we extend our treatment to even lower $T_{\rm eff}$ values.

\subsection{H lines}

CS is the version of the code that is implemented entirely in JS, and necessarily has a very limited 
line list of ~20 lines, and only included the Balmer series lines of \ion{H}{1} up to H$\epsilon$.  
Now that the SED is computed and displayed for $\lambda < 364$ nm, we have added two additional 
H lines H(2-8) and H(2-9), and a user who has their own installation can uncomment an additional
eight H lines, up to H(2-17).  The H line treatment includes Stark broadening, and these higher
Balmer lines allow for a somewhat more realistic treatment of the SED for $\lambda\ge$ 364 nm. 

\subsection{Scalable Vector Graphics}

As of our last report, CS used the HTML5 $<$canvas$>$ element for the graphical output.
This is not scale invariant, and the alphanumeric graphical elements were not sharp at any zoom
setting, and became increasingly pixelated in appearance at the higher zoom setting sometimes required 
for accurate lab work.  We now use the HTML5 $<$SVG$>$ (scalable vector graphics) element for the output
and the graphical elements now remain sharp at all zoom settings.  Moreover, the $<$SVG$>$ element
allows for interactivity based on event handlers to be added to the graphics, and we have taken advantage
of this to provide the UI with additional functionality that will help make the application more 
enticing at a basic level of pedagogy and outreach:

\begin{itemize}

\item{On all plots, when the user hovers, the data coordinates are displayed, and this allows for more
precise quantitative information to be extracted from the plots.}

\item{The user can now set the input stellar parameters by clicking on the HR diagram.}

\item{The user can now tune the narrow band filter by clicking on the rendering of the spectral image.}

\end{itemize}

\section{Discussion}

CSPy fills a gap between the research-grade stellar atmosphere and spectrum synthesis codes that
are compiled from fortran and require a unix-like environment, and the web-browser based
pedagogical modeling of CS, CSServ and CSDB.  Because python has a well developed set of support
tools such as IDEs and signal processing and plotting libraries, CSPy is a unique lab for 
studying and developing an astrophysical modeling code at the senior undergraduate or introductory
graduate level in an interactive 
and graphical way on common student-owned devices.

\paragraph{}

Python supports multi-threaded programing, and many commonplace devices now have multi-core CPUs,
so the way is open for improving the performance
of CSPy and other modeling codes in python, and enabling more realistic modeling in a 
pedagogically engaging environment.  Now that the code has been ported to python, it would
be relatively straightforward to port it to Julia, a language with similar syntax that
has been receiving increasing attention recently.  Julia is also an interpreted language, 
and so also offers the flexibility and transparency of an interpreted development and run-time
environment, but promises to yield executable code that runs significantly faster than that of python.  
Execution
speed is one of the main advantages that compiled languages like fortran have over interpreted
languages, so a port of ChromaStarPy to Julia could be significant.     

\paragraph{}

More generally, the OpenStars project is based on the philosophy that if it's worth computationally modeling  an
astronomical object for research purposes, then it's also worth using the model to render what the
object looks like in ways that people outside the research community, or who are learning the subject
at a more basic level, will find intuitive and relate-able, and can interact with.  Commonplace 
computing technology now allows for this, and this represents
a new way for the astronomy research and higher education community to be relevant beyond the 
research and higher education institutions.  
 We continue to encourage computational astrophysicists to consider didacticizing the modeling 
and visualization that they do and deploying it in forms that are relevant to education and
public outreach.



\acknowledgments
The author acknowledges Natural Sciences and Engineering Research Council of 
Canada (NSERC) grant RGPIN-2014-03979.  The author also thanks Saint Mary's
Astronomy graduate students Mitchell Young and Diego Casta\~neda for valuable 
guidance on python distributions.

\clearpage



\clearpage






\end{document}